# Ethical, Legal and Social aspects of Information and Communication Technology

Minati Mishra

Dept of ICT, FM University, Balasore

**Abstract:** In this era of computers and communication technology where computers and internet have made their ways to every sphere of life from offices to residences, reservation counters to banks to post offices, small retail shops to big organizations, health care units to entertainment industries etc., there emerged numerous questions regarding the ethical and legal uses of Information and Communication Technology (ICT). Like any other technological inventions ICT too has created both positive and negative impacts on the society. This paper aims at exploring some of these issues in brief.

## 1. Ethics and law

Ethics defines what is good for an individual as well as for the society and establishes the nature of duties that people owe themselves and one another. Human beings have the ability, partly innate and partly acquired, to judge human actions as morally good or bad, right or wrong. Even though "good / right" and "bad / wrong" do not mean the same thing for all still, everyone possesses a notion of right and wrong. Early morning music practice or use of insecticides/ repellents in her house, for example, though is very much ethical to my next-door neighbor; they are certainly not so to me! These differences again are not only individual but also cultural. Polygamy, for example, is a normal behavior and well acceptable to Muslims whereas not so for Hindus! Of course, in spite of all sorts of diversified opinions, ethics have a universal component. Killing an innocent person, for example, is not morally acceptable to anyone of us irrespective of our culture and belief.

Though law often embodies ethical principles, law and ethics are far from co-extensive. In an ideal world, the moral and legal acts may mean the same but they are not same in the real world. Many acts that would be widely condemned as unethical are not prohibited by law and the reverse is also equally true. Not helping one's friend when he/she is in need, for example, may not seems to be moral but not violation of any law too. Our state forbids recruitment of a non-reserved category candidate against few reserved category seats but, does not it contradict the overt moral imperative that there should be no discrimination among people based on cast, creed, religion and color?

Though it would be exaggerating and skeptical to infer that there is no connection altogether between ethics and the law but at most what the law will give us is a rough indication of practices, rather than an absolute criterion, valid for a particular place at a particular time deemed socially desirable, partly for moral reasons and partly for others. [1][2]

## 2. Ethical, Legal and Social aspects of ICT

Every technological invention has got both positive and negative impacts on the society. Einstein while giving the nuclear power theory as has never expected that his discovery shall ever be used for such a devastating destruction at Hiroshima and Nagasaki, it too was not known during the 19ths that communication technology of late shall have so many alarming direction associated with it.

For example, ICT provides easier and efficient means of storage and retrieval of information but at the same time suffers from piracy of copyrighted materials, software, data, music, video etc. at large scales. Internet provides instant access to all sorts of useful information at finger tip but at the same time suffers from plagiarism, illegal uploading, downloading, copying, stealing and misuse of intellectual property. ICT has created high-end job opportunities for the techies in one hand and on the other hand has created sever unemployment among non-tech groups. Communication Technology has made trade, investment, business simpler and unruffled through e-commerce and on-line transactions but suffers from cyber crimes, forgery, sabotage, hacking and loss. Internet has made the whole world a small intellectual village but at the same time is polluted with horrid contents like pornography, spam, worms and viruses. Therefore, it is high time now for careful inspection of the legal and ethical aspects of ICT as there are not enough guidelines available in this field as compared to those available in conventional branches of science and technology. More importantly, now ICT is not limited to the scientists and software engineers alone rather it has become a widespread phenomenon, affecting people at various stages in their role, as customers, service provider, participants, middlemen etc. So it has become the moral responsibility of the sociologist, business people and scientists to decide in which way ICT can be best utilized [8].

## 2.1 Privacy, Integrity, Security and Protection of Information & the Internet

*Privacy* means providing confidentiality to our personal data. One may not like, for example, to make public one's insurance details, medical history etc.

*Information Integrity* means the information provided should be relevant, complete, up-to-date, trustworthy and available in time. A job alert after the recruitment is over or without the address of the employer or to a student who is in his/her high school, for example, is hardly of any use.

*Security and protection* concerned with protection against accidental or intentional destruction or disclosure of data and programs by unscrupulous persons and in case data loss occurs how to recover it.

The Internet is a global technology network (WAN) made up of many smaller contributing networks (LANs, MANs, CANs etc.) to support the open exchange of information among many different kinds of institutions, organizations and individuals all over the world. Internet, no doubt, is a boon to the present society by providing answers to almost every problem and need of mankind but at the same time, can be its greatest enemy as well. We can be in touch with our friends thousands of miles away from us; can pursue a course in MIT while sitting in a small village of India, Consult with doctors of international repute sitting at our own home, reserve air/rail/bus tickets, sale/purchase goods, transact money, refer books/literatures/journals/newspapers and gather information with ease and almost free of cost and instantly which otherwise would have cost a lot of time, money and hectic travels. The above list is just indicative and the detail list can include lot many other usage. But unfortunately, in this age of Internet, it is very easy for scrupulous persons for attacking into one's privacy, to have access into one's private information, corrupt and destroy valuable data and assets though viruses and worms. And most importantly, the way day by day we are becoming more and more dependant on data processed by computers if, the input data is corrupted or there is some undetectable bug in the program then the output information is bound to lose integrity. Therefore utmost care should be taken maintain data integrity and unfortunately, the bad boys in the cyber world too are well aware of this fact and they work even harder to breach the security. And more unfortunately, the cyber crimes today are again not limited to the categories mentioned above but everyday there emerging some new forms of crimes. According to Dayanidhi Maran, the union minister of ICT, "The rapid increase of the computers and internet has given rise to new forms of crimes like – phishing and identity theft". "The banks are in a tough

[68]

position" says one FBI agent on context of the most brazen identity thief in US who has stolen more than $260 million of other people's money. According to him, "The Banks are torn between customer service and security. They want to make it easy for customers to access information, to see/ handle their accounts online. But that also makes it easy for the criminals". Abraham Abdallah, the audacious criminal has just took advantage of the loopholes in the most sophisticated on-line transaction system of US to gather the whole lot of money from other accounts before got caught by the New York Police Department (NYPD) in march 2001[3][4]. Though there is no official count of Indian victims of Identity Theft, the crime is steadily growing and according to RW Chua of online Fraud Asia which records and tries to eliminate online crime in Asian region, out of 50 Asians who go online at least one becomes a victim of identity crime. In US the estimated number of crimes in as large as 10 million per year and the crime is labeled by the Govt. authorities as the white-collar crime since late 1990s. And seeing the elevating identity theft rates in India, the IT Act of 2000 is being tightened. According to IT amendment Bill 2006, identity theft is made an offence punishable with up to five years of imprisonment and a fine. Another most common type of cyber scam of today is phishing in which the crooks send bogus emails tricking the user into giving up personal information at fake websites that resemble those of legitimate financial institutions and other commercial outfits. And the volume of these phishing emails is almost .7 billion messages per day, according to Symantec. Most probably all most all of us, who are using Internet for sometimes back, might have some experience of phishing emails. One afternoon during This June when I got a call at my residence from one of my colleague Dr. S Goswamy narrating how he is not allowed to operate his own Yahoo ID following certain correspondence with the Yahoo-group itself, I had experienced how crooked and clever these phishermen are. Another common type of scam is "pharming", in which genuine websites are hit with malicious codes that takes those visit them to look alike sites where every information given by the user is recorded without the sliest knowledge of the naïve[5][6]. This list of cyber crimes nowadays is just endless and therefore it is advisable first and fore mostly, to think carefully about what to say and how to say while communicating over the net as while communicating this way we don't see the physical persons we are communicating with, are not sure about their age, sex etc and there is no way to be assured of fact that what they are telling is true. Secondly, every user who goes on-line should have installed with a licensed security software and should update the database regularly

so as to enable the software to recognize the new signatures. Thirdly, unsolicited emails should always be deleted instantly without opening and firewalls, filters, security options should be set properly so as to block such unwanted and malicious traffic.

**2.2 ICT in Medicine and Health care**
ICT plays an increasingly important role in the field of medicines and diagnosis. Today, ICT in medicine is not only used for storing, processing, modeling and transmitting patient data rather has given rise to completely new diagnostic tools such as CT(computer tomography), CAT scan, ultrasound, MRI (magnetic resonance imaging), functional MRI, PET (positron emission tomography) etc that help inspection of the inner body without dissection or invasion. The new medical imaging methodologies where are opening enormous possibilities for diagnosis and scientific investigation at the same time are posing new epistemological, ethical and validity problems e.g., bodily properties that can be visualized in a one-to-one scale are emphasized in favor of those which cannot be locally and distinguishably represented within a picture, the pictures may contain artifacts stemming from technology itself or from the interaction between the technical depiction and the living body [7].

**3. Summary**
With the growth of ICT from its infantry of $19^{th}$ centuries to today's mighty form there emerged numerous questions regarding the ethical and legal uses of ICT from time to time. Ethics of ICT and indeed the ethical theory in general has typically been concerned with the normative analysis of individual intentional action. The standard topics of computer ethics such as invasions of privacy, misuse of personal data and disputed ownership of intellectual property appear to lend themselves well to such analysis, and because organizations and states are intentional agents too issues of regulation, legislation, corporate responsibility and other forms of collective action have also been relatively amenable to this approach. However, another set of normative issues exists of equal or arguably perhaps greater importance is the role played by ICT in globalization and eradication of poverty etc. Every technological developments and inventions, after all, are for the need and betterment of the society and should be helpful in the upliftment of the society.